\documentclass[prb,twocolumn]{revtex4}

\usepackage[dvips]{graphicx}
\usepackage[dvips]{color}
\usepackage{hyperref}

\usepackage{float}
\usepackage{amsmath}
\usepackage{amsfonts}

\newcommand{\abs}[1]{\left\lvert#1\right\rvert}

\newcommand{\mmax}[2]{\ifdim#1>#2#1\else#2\fi}
\newcommand{\mmin}[2]{\ifdim#1<#2#1\else#2\fi}

\begin{document}

\title{Spin-Orbit Coupling in Diamond and Zincblende Heterostructures}
\author{Miguel A. Oliveira}
\altaffiliation[Present Address: ]{Department of Physics, Trinity College,
Dublin 1, Ireland}
\thanks{Research supported by the FCT (Portugal) under grant PRAXIS XXI/BD/13349/97.}
\email{m.a.oliveira@tcd.ie}
\author{Angus MacKinnon}
\affiliation{Blackett Laboratory, Imperial College London, South Kensington Campus, London SW7 2AZ, UK}
\thanks{The authors wish to acknowledge the contribution of the late Prof. E. A.
        Johnson to this work.}
\date{\today}

\message{\the\linewidth}
%%%%%%%%%%%%%%%%%%%%%%%%%%%%%%%%%%%%%%%%%%%%%%%%%%%%%%%%
%
% Abstract:
%
%%%%%%%%%%%%%%%%%%%%%%%%%%%%%%%%%%%%%%%%%%%%%%%%%%%%%%%%

%=======================================================
                                         
\begin{abstract}
Spin splittings in III-V materials and heterostructures are of interest
because of potential applications, mainly in spintronic devices.
A necessary condition for the existence of these spin splittings is the
absence of inversion symmetry. In bulk zincblende materials the inversion
symmetry is broken, giving rise to a small spin splitting. The much larger
spin splitting observed in quantum wells is normally attributed to the
asymmetry of the confining potential and explained on the basis of the
Rashba effect. For symmetrically confined wells, where the only source of
asymmetry is that of the underlying crystal potential, the confining potential
strongly enhances the spin splittings. This enhancement does not require the
asymmetry of the confining potential but depends on the interplay between the
confinement and the crystal potential. In this situation the behavior of the
spin splittings is consistent with the Dresselhaus contribution.

In asymmetrically confined wells both Dresselhaus and Rashba terms contribute.

We present a general theory of the spin splittings of these structures based on
the group theory of diamond and zincblende heterostructures.
\end{abstract}

%=======================================================

%%%%%%%%%%%%%%%%%%%%%%%%%%%%%%%%%%%%%%%%%%%%%%%%%%%%%%%%

\maketitle
%%%%%%%%%%%%%%%%%%%%%%%%%%%%%%%%%%%%%%%%%%%%%%%%%%%%%%%%
%
% Section I:
%
%%%%%%%%%%%%%%%%%%%%%%%%%%%%%%%%%%%%%%%%%%%%%%%%%%%%%%%%

%=======================================================
                                         
\section{Introduction}
                                                
%=======================================================

\label{intro}

Spin is one of the most intriguing properties of subatomic particles and its
explanation is among the most significant achivements of quantum theory.
Although the field of electronics has taken full advantage of electronic
charge, spin has been relatively unexploited in any practical application with
the noteable exception of magnetic read heads based on giant
magnetoresistance.
However recent developments suggest that this is about to change.
The new field of spintronics \cite{awschalom} is steadily growing with the aim
of taking full advantage of spin as well as charge. Some devices
have been already proposed \cite{datta,hall} and some experimental
work \cite{murdin} is already taking advantage of some basic spin
properties of heterostructures.

However the spin properties of heterostructures of IV and
III-V materials are still not completely understood. In spite of recent
advances,\cite{cartoixa1,golub} no detailed atomistic
simulations are yet available (to our knowledge) on the spin properties of the
materials that form the backbone of the semiconductor industry.
Most of the work reported in the literature relates to
a particular form of the $\vec{k}\cdot\vec{p}$ model.\cite{cartoixa1,golub}
It is only recently that this model has been able to incorporate
the full symmetries of these heterostructures.\cite{cartoixa1}

We shall present a method of predicting the spin splittings of these
structures and we shall present results based on the Empirical
Pseudopotential Layer Method (EPLM) that corroborate our claims. The
often discussed contributions of bulk inversion asymmetry (BIA) and
structural inversion asymmetry (SIA) will be presented as consequences of
the symmetries of the heterostructure and we shall show clearly that any
detailed calculation must include both.

Section~\ref{methods} will contain a summary description of the computational
methods. Section~\ref{symmetries} will focus on the consequences of symmetry
for the spin splittings of these structures. The following sections contain a
discussion of our results followed by section~\ref{concl} where conclusions
will be drawn.

%=======================================================

%%%%%%%%%%%%%%%%%%%%%%%%%%%%%%%%%%%%%%%%%%%%%%%%%%%%%%%%

%%%%%%%%%%%%%%%%%%%%%%%%%%%%%%%%%%%%%%%%%%%%%%%%%%%%%%%%
%
% Section II:
%
%%%%%%%%%%%%%%%%%%%%%%%%%%%%%%%%%%%%%%%%%%%%%%%%%%%%%%%%

%=======================================================
                                         
\section{The computational models}
                                                
%=======================================================

\label{methods}

The EPLM has already been discussed elsewhere \cite{rundell1} so
we shall only describe it here briefly, mainly to point out the
particulars of our implementation. 

In a conventional band structure calculation a set of eigenenergies is
calculated for a particular value of the wavevector, $\vec{k}$.  By contrast a
scattering approach, such as EPLM, works with a fixed energy and
$\vec{k}_\parallel$
parallel to an interface and calculates a set of solutions for $k_\perp$
perpendicular to that interface.  These solutions include examples with
real and complex $k_\perp$, often called the {\it complex\/} band
structure.  Solutions for different layers may be combined using appropriate
matching conditions to generate solutions for more complicated combinations of
layers.  Typically the result may be expressed as a transmission coeficient
for the multi--layer system.  Eigenstates of the system then manifest
themselves as resonances in the transmission coefficient or as bound states
decaying into the gap of the embedding material. 

We also consider a much simpler model in which the system is embedded in an
infinite well. 

%=======================================================
                                                                                                                        
\subsection{The EPM and the matching conditions}
                                                                                                                        
%=======================================================

The first step of the method is to compute the complex band structure
in each layer of the heterostructure using the Schr\"odinger
equation in $\vec{k}$--space:
\begin{eqnarray}
H_{\vec{G},\vec{G}'}^{s,s'}&=&\left[\frac{\hbar^2}{2m}(\vec{k}+\vec{G})^2-E\right]%
\delta_{\vec{G},\vec{G}'}\delta_{s,s'}\nonumber\\
&&+V(\vec{G}-\vec{G}')\delta_{s,s'}%
-I_{\vec{G},\vec{G}'}^{s,s'},
\label{xxx}\end{eqnarray}
where $I_{\vec{G},\vec{G}'}^{s,s'}$ represents the spin-orbit term.

The crystal potential is treated as a local pseudo\-potential \cite{cohenbook}
described in terms of atomic form factors:
\begin{equation}
V(\vec{G}\,)=\sum\limits_{\alpha}v_{\alpha}(\vec{G}\,)S_{\alpha}(\vec{G}\,),
\end{equation}
where $S_{\alpha}(\vec{G}\,)$ represents the structure factor and
$v_{\alpha}(\vec{G}\,)$ the form factors for atom species $\alpha$.

The spin-orbit term is included in its usual for\-mu\-la\-tion:
\cite{weiszarticle,bloomarticle1,bloomarticle2}
\begin{eqnarray}
I_{\vec{G},\vec{G}'}^{s,s'} & = & -i\sum\limits_{\alpha}\lambda_\alpha%
S_{\alpha}(\vec{G}-\vec{G}') \nonumber\\%
&&\times \left[(\vec{k}-\vec{G})\wedge(\vec{k}-\vec{G}')\right]\cdot%
\vec{\sigma}_{s,s'}\nonumber\\
& = & -i\Lambda(\vec{G}-\vec{G}')%
\left[(\vec{k}-\vec{G})\wedge(\vec{k}-\vec{G}')\right]\cdot%
\vec{\sigma}_{s,s'}, 
\end{eqnarray}
where $\vec{\sigma}$ represents the usual vector of Pauli matrices and
$\lambda_\alpha$ the spin-orbit form factors.

In the case of heterostructures the symmetry is retained in the plane
perpendicular to the growth direction and the component of the wave vector
$\vec{k}$ parallel to this plane is still a good quantum number. $\vec{k}$ may
be decomposed into $\mbox{$\vec{k}=(\vec{k}_\parallel,k_\perp)$}$. 
The Hamiltonian can be written as a polynomial expansion in $k_\perp$:
\begin{equation}
H_{\vec{G},\vec{G}'}^{s,s'}=H_{2,\vec{G},\vec{G}'}^{s,s'}k_\perp^2+%
H_{1,\vec{G},\vec{G}'}^{s,s'}k_\perp+H_{0,\vec{G},\vec{G}'}^{s,s'},
\end{equation}
where:
\begin{eqnarray}
H_{2,\vec{G},\vec{G}'}^{s,s'} & = & \frac{\hbar^2}{2m}\delta_{\vec{G}',\vec{G}}%
\delta_{s',s},\nonumber\\
H_{1,\vec{G},\vec{G}'}^{s,s'} & = & \frac{\hbar^2}{m}G_\perp\delta_{\vec{G}',\vec{G}}%
\delta_{s',s}-i\Lambda(\vec{G}-\vec{G}')\vec{A}\cdot%
\vec{\sigma}_{s',s},\nonumber\\
H_{0,\vec{G},\vec{G}'}^{s,s'} & = & \left[\frac{\hbar^2}{2m}(\vec{k}_\parallel+\vec{G})^2-E\right]%
\delta_{\vec{G}',\vec{G}}\delta_{s',s}\nonumber\\%
&&+V(\vec{G}-\vec{G}')\delta_{s',s}
 -i\Lambda(\vec{G}-\vec{G}')\vec{B}\cdot\vec{\sigma}_{s',s},
\end{eqnarray}
where we have used the definitions:
\begin{eqnarray}
\vec{A}
& = &
\hat{k}_\perp\wedge(\vec{G}-\vec{G}'),\\
\vec{B}
& = &
\vec{k}_\parallel\wedge(\vec{G}-\vec{G}')+\vec{G}'\wedge\vec{G}.
\end{eqnarray}
It can be shown \cite{pendry,chang} that this equation 
may be recast as an eigenvalue problem in $k_\perp$ for fixed energy,
$k_\parallel$ and growth direction.

This eigenproblem gives all the required information, namely all the $k_\perp$
and the corresponding eigenvectors, to allow the wavefunction
to be completely determined.

If a complex band structure is determined for adjacent layers $i$ and $i+1$,
with an appropriate band offset, regular matching conditions can be imposed
as:
\begin{eqnarray}
\Psi_i(\vec{r}\,) & = & \Psi_{i+1}(\vec{r}\,),\\
\frac{\partial \Psi_i}{\partial z}(\vec{r}\,) & = & \frac{\partial \Psi_{i+1}}{\partial z}(\vec{r}\,),
\end{eqnarray}
at the interface between layers. These conditions can be re-expressed as matrix
conditions connecting the wave functions in both layers. A predetermined wavefunction
in the first layer results then in fixed coefficients across the structure.

%=======================================================
                                                                                                                        
\subsection{The infinite well models}
                                                                                                                        
%=======================================================

Using the matching conditions and the complex band structure information from
last section it is simple to fix infinite well conditions at the extremities
of the structure. Denoting by $0$ the left interface for the first layer and
by $N$ the right interface for the end layer we shall have:
\begin{equation}
\Psi_0=\Psi_N=0.
\end{equation}

These conditions in conjunction with the matching conditions form a set of
equations whose solution is usually expressed as a determinant.\cite{brand}
Determining the solution is however best tackled by singular value
decomposition techniques. With this approach it is easy to determine all
the energy levels for the system by analyzing the behavior of the singular
values.

It should nevertheless be mentioned that the method still suffers from all 
the problems described previously.\cite{brand}

%=======================================================
                                                                                                                        
\subsection{The Empirical Pseudopotential Layer Method}
                                                                                                                        
%=======================================================

The EPLM is far more general than the infinite well models.
The numerical problems inherent in that method are not
present and appropriate boundary offsets and materials can
be selected.

Using the complex band structure information and the matching
techniques a scattering matrix approach may be implemented.\cite{ko1,ko2}
The energy levels are determined by analyzing the resonances of the
transmission coefficient across the structure. By calculating
the wavefunctions at those energies all properties are then
accessible.

This method is extremely well suited to the study of general
heterostructures as no assumptions need be made about its layout or
its growth direction. The band offsets between layers are
taken from experimental values.

%=======================================================
                                                                                                                        
\subsection{Wavefunction based calculations}
                                                                                                                        
%=======================================================

Both methods give enough information, after an initial energy level
determination, to compute the wavefunction or any other observables.
In fact for a fixed growth direction, energy and parallel wavevector
the methods supply a complete description of the wavefunctions.
This information is then used to compute the relevant properties.

In particular we compute the parallel averaged probability density,
given by:
\begin{equation}
\rho(r_\perp)=\int\ d^2r_\parallel \Psi^\dagger(\vec{r}\,)%
\Psi(\vec{r}\,),
\end{equation}
which can then be used to compute the total probability density:
\begin{equation}
P=\int\ dr_\perp\sigma(r_\perp).
\end{equation}

Another useful quantity is the parallel averaged spin polarization
given by:
\begin{equation}
\bar{\sigma}_i(r_\perp)=\int\ d^2r_\parallel \Psi^\dagger(\vec{r}\,)%
\sigma_i\Psi(\vec{r}\,),
\end{equation}
which will then give a total spin polarization of the form:
\begin{equation}
\bar{\sigma}_i=\frac{\int\ dr_\perp\bar{\sigma}_i(r_\perp)}{P}.
\end{equation}

The method is flexible enough to compute any other relevant observable if
necessary.

%=======================================================

%%%%%%%%%%%%%%%%%%%%%%%%%%%%%%%%%%%%%%%%%%%%%%%%%%%%%%%%

%%%%%%%%%%%%%%%%%%%%%%%%%%%%%%%%%%%%%%%%%%%%%%%%%%%%%%%%
%
% Section III:
%
%%%%%%%%%%%%%%%%%%%%%%%%%%%%%%%%%%%%%%%%%%%%%%%%%%%%%%%%

%=======================================================

\section{Symmetries}

%=======================================================

\label{symmetries}

%=======================================================
                                         
\subsection{Basic definitions}
                                                
%=======================================================

Let us start by firmly setting the scope of our work. We are interested in
the spin physics of lattice matched heterostructures of diamond--like and
zincblende materials.

In the possible plethora of all these structures it is useful to separate
them into categories. Firstly we consider the atomic layer layout. If
these structures have a mirror symmetric atomic layer distribution we
say it is a symmetric structure. Examples of these are often used and include,
for instance, layouts of GaAs in AlAs.

We have to stress that although the atomic layer layout may be symmetric
the atomic positions within the layers are such that the layers are not
strict mirror images of each other.

Any layout that is not symmetric is said to be asymmetric.

Another useful classification considers the sharing or not of a common anion
in the structure. The case of GaAs in AlAs is a clear case of a common--anion 
structure. There are however situations where this does not happen.
For example, a heterojunction of GaSb and InAs is such a case.
This is then said to be a no--common--anion structure.

These definitions will later become important in characterizing the symmetries
of the structures.

%=======================================================
                                         
\subsection{Symmetries}
                                                
%=======================================================

For bulk semiconductors the spin splittings are determined by the symmetry
of the crystal lattice. Diamond has point group $O_\mathrm{h}$ and zincblende
$T_\mathrm{d}$.
These determine which terms are allowed in the Hamiltonian, whether
the spin splittings are possible and, in that case, which form they have.

In the case of heterostructures the symmetry is reduced and it is important
to know which subgroup of the bulk group a particular structure has.
It is however impractical to enumerate all the point groups for every
possible layout and every possible growth direction. We confine ourselves
therefore to the most common cases.

In the case of diamond--like materials the most
usual layouts consist of layers of Si and Ge grown in the [001] direction.
In symmetric configurations these structures have either point group
$D_\mathrm{2d}$ or $D_\mathrm{2h}$. An odd number of atomic layers of one embedded in
the other has point group $D_\mathrm{2d}$ while an even number has point group
$D_\mathrm{2h}$.

Non ideal interfaces containing monatomic fluctuations can also produce
structures with point groups $C_\mathrm{2v}$, $C_\mathrm{4v}$ and
$D_\mathrm{4h}$.\cite{golub} We shall however not consider these cases as the
methods used in this work can only handle perfect interfaces.

In the case of zincblende heterostructures with a common anion grown
in the [100] direction we conclude that symmetric structures have
point group $D_\mathrm{2d}$ while asymmetric structures have point
group $C_\mathrm{2v}$. For structures without a common anion we
have point group $C_\mathrm{2v}$.

A summary of these cases is given in table~\ref{spg}.

\begin{table}[H]
\begin{center}
\begin{tabular}{|c|c|c|c|}
\hline
Diamond    & Symmetric  & odd number of atomic layers  & $D_\mathrm{2d}$ \\
\cline{3-4}
  [001]    &            & even number of atomic layers & $D_\mathrm{2h}$ \\
\cline{2-4}
           & Asymmetric &                              & $C_\mathrm{2v}$ \\
\hline
Zincblende & Symmetric  & Common-anion                 & $D_\mathrm{2d}$ \\
\cline{3-4}
  [001]    &            & No common-anion              & $C_\mathrm{2v}$ \\
\cline{2-4}
           & Asymmetric &                              & $C_\mathrm{2v}$ \\
\hline
\end{tabular}
\end{center}
\caption[Summary of point groups for heterostructures.]
{\label{spg}Summary of point groups for heterostructures
grown in the [001] direction.}
\end{table}

However, these considerations only provide us with a rule--of--thumb.
The point group of a particular heterostructure must be determined for
that particular case. Many arrangements with only slight alterations
can be produced which have different point groups.  For other growth
directions similar considerations apply but the resulting point groups will
generally be different.

%=======================================================
                                         
\subsection{Symmetries and the Hamiltonian}
                                                
%=======================================================

In any particular situation we can always consider the
Hamiltonian as an expansion in powers of $\vec{k}_\parallel$ about a high
symmetry point in the 2D Brillouin zone, usually $\Gamma$.  Indeed this is
the basis of the popular $\vec{k}\cdot\vec{p}$ approximation.
The particular terms we are interested in are those involving
spin of the form:
\begin{equation}
\label{sym_terms}
\gamma_{\alpha,\beta,\cdots,\zeta,\cdots}\sigma_\alpha k_\beta\cdots k_\zeta\cdots,
\end{equation}
where $\gamma_{\alpha,\beta,\cdots,\zeta,\cdots}$ is a case dependent constant, 
$\sigma_\alpha$ one of the Pauli matrices and the $k_\beta\cdots k_\zeta\cdots$
a product of components of the $\vec{k}_\parallel$ vector. 

Most of these terms are not allowed by symmetry and may be excluded.
In fact, when we are interested in the behavior near an extremum such as the $\Gamma$ point, it is
usually sufficient to consider the first few terms.

In the case of diamond--like structures with point group $D_\mathrm{2h}$ no spin-orbit terms
of the form (\ref{sym_terms}) are allowed in the Hamiltonian and hence no spin
splittings should be observed. This is easily understood as this point group has as a
constituent symmetry the inversion center.

In the case of the point group $D_\mathrm{2d}$ this is however not the case. Linear terms
like:
\begin{equation}
\sigma_x k_x-\sigma_y k_y,
\end{equation}
are possible and thus linear splittings may be observed.

This linear contribution can be understood in terms of the cubic
terms \cite{andrada1,eppenga,yakonov} in the bulk
and has hence been coined the Dresselhaus or bulk inversion
asymmetry (BIA) term.

\begin{figure}[H]
\begin{center}
\includegraphics[width=0.8\linewidth]{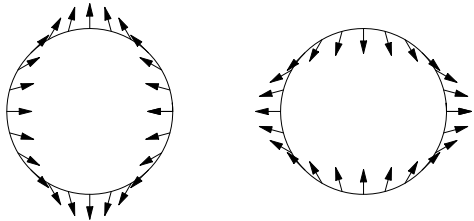}
\end{center}
\caption[Spin diagrams for the Dresselhaus contribution.]
{\label{sd_dressel}Spin diagrams for the Dresselhaus
contribution. Magnitude of spin scaled for clarity.}
\end{figure}

A simple toy model can be constructed with this term and it is
easy to calculate the spin
polarization as a function of $\vec{k}_\parallel$. This dependence
for a fixed magnitude of $\vec{k}_\parallel$ plotted at regular
angular intervals will henceforth be called a spin diagram. For
this particular case the two possible  spin diagrams are displayed
in figure~\ref{sd_dressel}.

In the case of a structure with $C_\mathrm{4v}$ the only invariant that
can be found is the term:
\begin{equation}
\sigma_x k_y-\sigma_y k_x,
\end{equation}
which again allows for linear splittings. This contribution
was understood early by
Rashba \cite{rashba1,rashba2} in terms of the consequences
of structural asymmetry in the material. This term is usually
called the Rashba or structural inversion asymmetry (SIA)
term.

\begin{figure}[H]
\begin{center}
\includegraphics[width=0.8\linewidth]{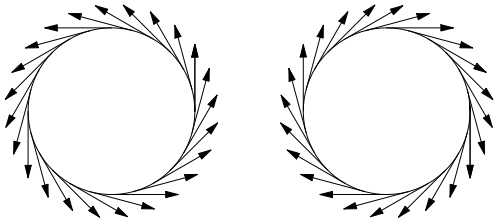}
\end{center}
\caption[Spin diagrams for the Rashba contribution.]
{\label{sd_rashba}Spin diagrams for the Rashba
contribution. Magnitude of spin scaled for clarity.}
\end{figure}

The toy model can be repeated with the Rashba term and its
characteristic spin diagram is depicted in figure~\ref{sd_rashba}.

The determination of the spin diagrams is an easy method to
visualize the symmetries as these act as symmetry signatures.

\begin{figure}[H]
\begin{center}
\includegraphics[width=0.8\linewidth]{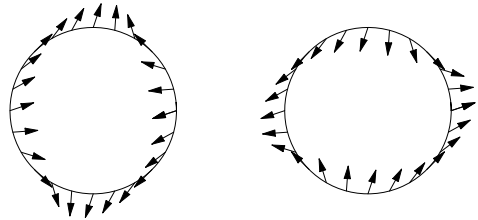}
\end{center}
\caption[Spin diagrams for mixed BIA and SIA contributions.]
{\label{sd_dr}Spin diagrams for a particular case of mixed BIA and
SIA contributions. Magnitude of spin scaled for clarity.}
\end{figure}

For point group $C_\mathrm{2v}$ both BIA and SIA terms are allowed
and the spin diagram looks like a superposition of both.
An example is shown in figure~\ref{sd_dr}.

In this situation the angle of the spin
direction at $\vec{k}_\parallel=(k_x,0)$ with the [100] direction
is a good measure of the degree of mixing.

The zero--field spin splittings may be represented
\cite{andrada1} by an effective magnetic
field. This will produce a spin Hamiltonian term of the form:
\begin{equation}
H=\frac{1}{2}\hbar\vec{\sigma}\cdot\vec{B}_\mathrm{eff}(\vec{k}_\parallel\,),
\end{equation}
where $\vec{B}_{\mathrm{eff}}(\vec{k}_\parallel\,)$ will depend on the magnitude
and direction of the parallel wavevector $\vec{k}_\parallel$.
 From this
Hamiltonian we obtain an overall spin splitting given by:
\begin{equation}
\Delta(\vec{k}_\parallel)=\hbar\abs{\vec{B}_\mathrm{eff}}.
\end{equation}

From our previous discussion we can already deduce that $\vec{B}_{\mathrm{eff}}$
will have two contributions: the bulk term (BIA) and a structural
term (SIA).

In the case of bulk zincblende structures this term
produces a well known contribution \cite{andrada1,yakonov}
for small values of $\vec{k}$ of the form:
\begin{equation}
\label{cubicterms}
\vec{B}_{\textrm{eff}}=\frac{2\gamma}{\hbar}\left[k_x(k_y^2-k_z^2)\hat{x}+%
k_y(k_z^2-k_x^2)\hat{y}+k_z(k_x^2-k_y^2)\hat{z}\right]
\end{equation}
where $\gamma$ is a material dependent constant. In the case of our
structures this has been shown \cite{eppenga,andrada1}
to simplify to the form:
\begin{equation}
\vec{B}_\mathrm{BIA}=\frac{2\gamma}{\hbar}(k_{z,w}^2)(-k_x\hat{x}+k_y\hat{y}),
\end{equation}
where $k_{z,w}$ is the value of the confined wavevector in the well.
This term is exactly the one predicted by symmetry and will produce
a spin splitting $\Delta_{\mathrm{BIA}}$, that is linear in $\vec{k}_\parallel$
and isotropic.

In the case of structural inversion asymmetry it was
shown \cite{rashba1,rashba2} that this
effective field is:
\begin{equation}
\vec{B}_{\mathrm{R}}=\frac{\alpha}{\hbar}(\vec{k}\wedge\hat{k}_\perp),
\end{equation}
where $\hat{k}_\perp$ is just the unit vector in the growth direction.
For the particular case of the [001] growth direction this is then:
\begin{equation}
\vec{B}_{\mathrm{R}}=\frac{\alpha}{\hbar}(k_y\hat{x}-k_x\hat{y}).
\end{equation}
The spin splitting for this case $\Delta_\mathrm{R}$, is again linear in $\vec{k}_\parallel$ and also isotropic.

In the general case where both terms can co-exist a total spin splitting is given by:
\begin{equation}
\Delta(\vec{k}_\parallel)=\hbar\abs{\vec{B}_{\mathrm{BIA}}+\vec{B}_\mathrm{R}},
\end{equation}
which can be expressed as:
\begin{equation}
\Delta(\vec{k}_\parallel)=\sqrt{\Delta_{\mathrm{BIA}}^2+\Delta_R^2-2\Delta_{\mathrm{BIA}}\Delta_R\sin(2\theta)},
\end{equation}
where $\theta$ is the angle between $\vec{k_\parallel}$ and the [100] direction.
In general this spin splitting is linear in $\vec{k}_\parallel$
but anisotropic.

%=======================================================
                                         
\subsection{Symmetries and the simulation methods}
                                                
%=======================================================

\label{sym_sim}

Both models used in this work, the infinite well models
and the Empirical Pseudopotential Layer Method, contain
atomistic information about the structure under consideration
and should hence reproduce the full symmetries of the cases
under study. There are however some practical but soluble
problems.

Firstly it should be noted that the method uses only integer numbers
of monolayers (\mbox{i.e.} pairs of atomic layers) which takes some
cases out of our reach. An obvious example is the single layer of Si
in Ge. As this restriction only comes about as a simplification
in the matching technique it is possible to remove it if any of
the cases in this category becomes important.

Another problem concerns simulations with common atoms across
different material layers.
We should remember that the Empirical Pseudopotential
Method uses form factors determined for each material
individually and hence the common atom is described by
different potentials in different layers. This may cause
a further reduction in the symmetry which is
visible in the results but, as we shall see, does not
invalidate them. A change to consistent atomic form factors
would only partially solve this problem. Even if the same
form factors are used for the common atom, and because
the algorithm forces us to use integer number of monolayers,
we would have one atomic layer of these anions at a
different offset. However if we solve the matching at
integer number of atomic layers this problem would also be solved.

A few notes on the particular simulation cases chosen are
appropriate at this stage. As a simulation method the
Empirical Pseudopotential Method is quite robust and
powerful because the layout of the heterostructure is
completely arbitrary: the growth direction, the number of 
layers, which materials and which band offsets are all
set as input. This much freedom allows simulations with
structures whose layout is completely artificial. These
cases are however as important as those of naturally
occurring heterostructures. If the latter give us precious
data comparable with experimental results the former
enable us to explore every possible dependence
on the heterostructures' defining characteristics by carrying out computer
experiments which would not be possible in a real laboratory.

We should note that the artificial cases considered are
not so far from physical situations that render then
absurd. For example the infinite well situations are
attainable by using wide gap materials or even insulators
to confine the system. Varying band offsets is also
feasible to some extent by using alloying techniques.

%=======================================================

%%%%%%%%%%%%%%%%%%%%%%%%%%%%%%%%%%%%%%%%%%%%%%%%%%%%%%%%

%%%%%%%%%%%%%%%%%%%%%%%%%%%%%%%%%%%%%%%%%%%%%%%%%%%%%%%%
%
% Section IV:
%
%%%%%%%%%%%%%%%%%%%%%%%%%%%%%%%%%%%%%%%%%%%%%%%%%%%%%%%%

%=======================================================
                                         
\section{Diamond heterostructures}
                                                
%=======================================================

\begin{figure}[H]
\begin{center}
\includegraphics[width=\mmin{10cm}{0.8\linewidth}]{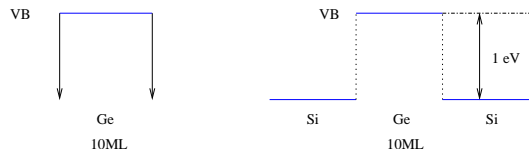}
\end{center}
\caption[Well layouts for diamond-like structures.]
{\label{layout1}Well layouts for diamond-like structures.
The layout on the left represents 10 monolayers of Ge in infinite
walls under the valence band while the right layout represents
10 monolayers of Ge in Si with an artificial offset to produce a
\mbox{1 eV} deep well adjusted on the valence band edges.}
\end{figure}

For the crucial case of diamond heterostructures we have used
two distinct simulations.
First we considered the case of a layer of Ge in infinite
walls. This structure should also have point group $D_{\mathrm{2h}}$
and no spin splittings should arise. The second case is that
of a layers of Ge in Si with an artificial offset to produce
a well for holes in the intermediate layer. This situation
also has point group $D_{\mathrm{2h}}$ and no spin splittings
should be observed. We should point out that strain effects
at the interfaces have been completely ignored. The two
physical situations are represented in figure~\ref{layout1}.

\begin{figure}[H]
\begin{center}
\includegraphics[height=\mmin{12cm}{\linewidth},%
width=\mmin{12cm}{0.8\linewidth},clip]{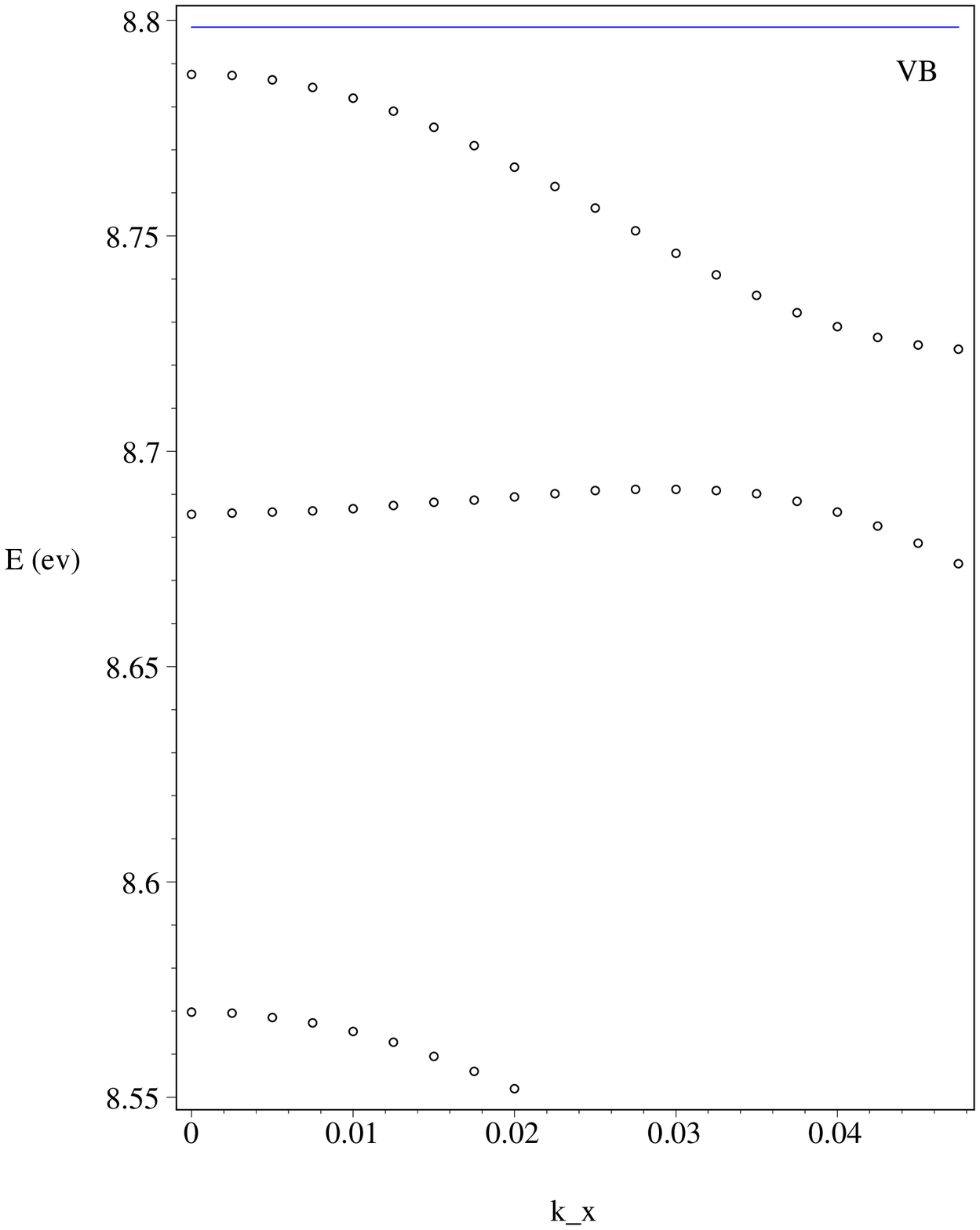}
\end{center}
\caption[Energy dispersion.]
{\label{well1}Energy dispersion for the well layout represented
on the left of figure~\ref{layout1}. $\vec{k}_\parallel=(k_x,0)$ in
units of $\left(\frac{2\pi}{a}\right)\,\textrm{\AA}^{-1}$. VB represents
the valence band edge of Ge.}
\end{figure}

\begin{figure}[H]
\begin{center}
\includegraphics[height=\mmin{12cm}{\linewidth},%
width=\mmin{12cm}{0.8\linewidth},clip]{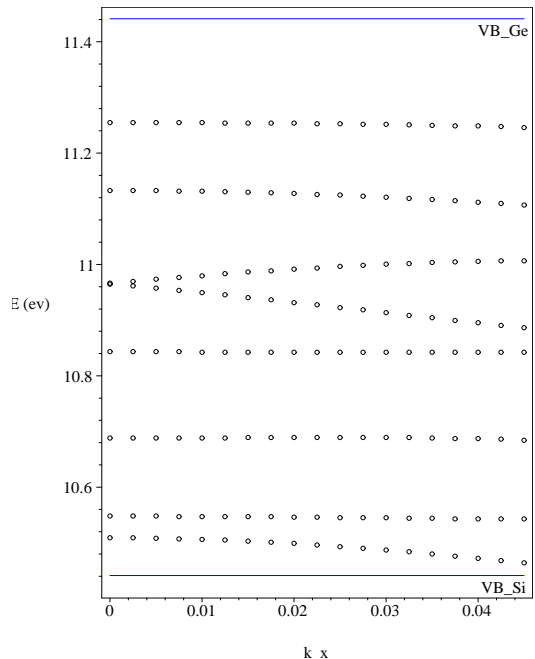}
\end{center}
\caption[Energy dispersion.]
{\label{well2}Energy dispersion for the layout represented
on the right of figure~\ref{layout1}. $\vec{k}_\parallel=(k_x,0)$ in
units of $\left(\frac{2\pi}{a}\right)\,\textrm{\AA}^{-1}$. $\mbox{VB}\_\mbox{Ge}$ represents
the valence band edge of Ge and $\mbox{VB}\_\mbox{Si}$ the valence band edge of Si.}
\end{figure}

Both simulations were run on extremely small energy grids and
the results showed no spin splittings at all as we can see in
figures~\ref{well1} and \ref{well2} for the two situa\-tions.
In the second situation it should be noted that the two very
close energy levels correspond to two twofold degenerate energy
levels and not to one energy level displaying spin splitting.
This was verified by a computation at the $\Gamma$ point were
the energy levels are close but not degenerate.

These results are hence in line with our predictions.

Some claims have been put forward \cite{golub}
that in principle we could engineer spin splittings by producing
a structure with a point group different from $D_{\mathrm{2h}}$. As we
have established, this corresponds to the case of an odd number of atomic
layers in the well. This situation is also extremely interesting:
although bulk diamond--like structures do not show any spin splittings,
and hence the Dresselhaus term cannot be present, in the case of
heterostructures this term is present. It is hence possible to have
spin splittings originating from a Dresselhaus contribution even if
the terms are absent in bulk material. There is then an alternative
way to engineer spin splittings in these structures which does not
rely on the Rashba effect. However this case is out of the reach of our
simulation methods in their current form.

\begin{figure}[H]
\begin{center}
\includegraphics[width=\mmin{10cm}{0.8\linewidth}]{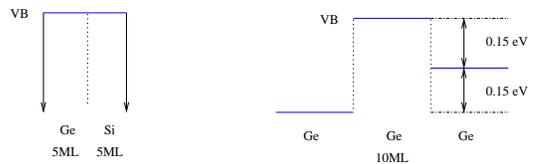}
\end{center}
\caption[Well layouts for diamond-like structures.]
{\label{layout2}Well layouts for diamond-like structures.
The layout on the left represents 5 monolayers of Ge and 5 monolayers
of Si in infinite walls with lined-up valence bands while the right
layout represents 10 monolayers of Ge in Ge with artificial
asymmetric offsets.}
\end{figure}

There are nevertheless other ways. To show that spin splittings
can indeed be achieved in structures involving diamond--like materials we tested
two situations that break inversion symmetry. The first is a sandwich of
two layers, one of Si and the other of Ge, in infinite walls with an
appropriate offset to line up the valence band edges. The second
consists in the artificial case of Ge sandwiched between layers of Ge but
with an artificial asymmetric offset. These cases do not have point group
$D_{\mathrm{2h}}$ and do not have an inversion center; spin splittings
are therefore allowed. Both situations are represented in figure~\ref{layout2}.

\begin{figure}[H]
\begin{center}
\includegraphics[width=\mmin{12cm}{0.8\linewidth}]{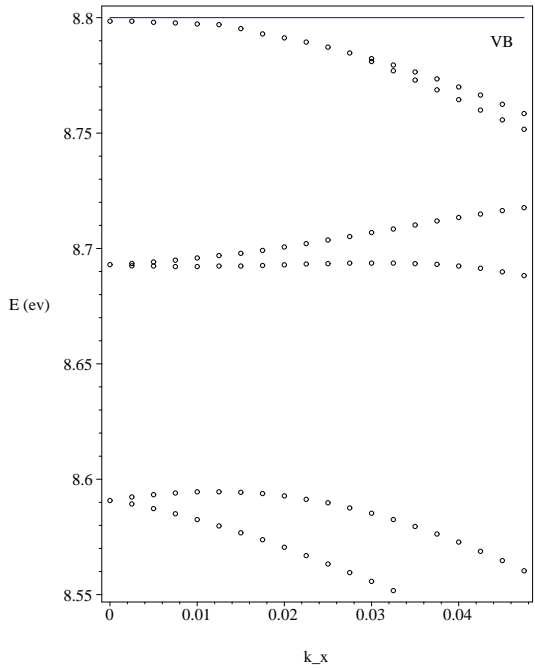}
\end{center}
\caption[Energy dispersion.]
{\label{well3}Energy dispersion for the well layout represented
on the left of figure~\ref{layout2}. $\vec{k}_\parallel=(k_x,0)$ in
units of $\left(\frac{2\pi}{a}\right)\,\textrm{\AA}^{-1}$. VB represents the
artificially set common valence band edge.}
\end{figure}

\begin{figure}[H]
\begin{center}
\includegraphics[width=\mmin{12cm}{0.8\linewidth}]{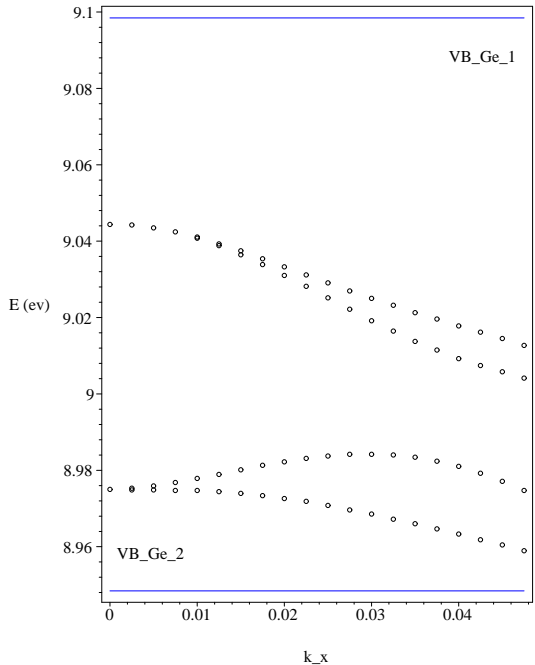}
\end{center}
\caption[Energy dispersion.]
{\label{well4}Energy dispersion for the well layout represented
on the right of figure~\ref{layout2}. $\vec{k}_\parallel=(k_x,0)$ in
units of $\left(\frac{2\pi}{a}\right)\,\textrm{\AA}^{-1}$. $\mbox{VB}\_\mbox{Ge}\_i$ represents
the artificially set valence band edges for the $i^\mathrm{th}$ layer.}
\end{figure}

The resulting energy dispersions for these two situations are
in figures~\ref{well3} and \ref{well4} respectively.

As we can see spin splittings are apparent in both situations
and hence these materials show potential for spintronic
devices.

The situation of infinite potential at the boundaries is also
a case of perfect symmetric confinement. In this situation no
extra asymmetry, other than the crystal potential, can cause the
spin splittings and only an interplay between crystal
potential and confinement is responsible for them. This case may
be described as spin splitting enhancement by symmetric
confinement and would have occurred even if a single zincblende
compound had been used, as we shall see later. This
enhancement refers to values of spin splitting far bigger
than those observed in bulk.

Two main conclusions may be drawn. Firstly, the two
simulation methods we have used are perfectly capable of
handling situations where by symmetry no spin splitting
is possible. Secondly, and more importantly, even in the case
of materials where a center of inversion is present structures
can be engineered which have spin splittings. This is of the
utmost importance as Si and Ge are presently the basis of
most of the semiconductor industry.

%=======================================================

%%%%%%%%%%%%%%%%%%%%%%%%%%%%%%%%%%%%%%%%%%%%%%%%%%%%%%%%

%%%%%%%%%%%%%%%%%%%%%%%%%%%%%%%%%%%%%%%%%%%%%%%%%%%%%%%%
%
% Section V:
%
%%%%%%%%%%%%%%%%%%%%%%%%%%%%%%%%%%%%%%%%%%%%%%%%%%%%%%%%

%=======================================================
                                         
\section{Zincblende common-anion heterostructures}
                                                
%=======================================================

The case of symmetric structures is still somewhat controversial in the literature.
Most of this controversy stems from the fact that the conventional model of
electronic structure in heterostructures, the $\vec{k}\cdot\vec{p}$ method,
does not fully account for their symmetries. In fact, in this method the
wavefunctions are expanded in a set of $\Gamma$ Bloch states of the
zincblende crystal. Further this expansion is restricted to a few states,
usually the top of the valence band and the bottom of the conduction band.
With this set it is impossible to resolve any atomistic details and the
method is incapable of reproducing the correct point group symmetries of
the structure. This was thought not to be problematic given the small values
of the bulk terms from BIA. However, work \cite{eppenga}
as early as 1988 hinted that this is not the case.
More recent theoretical studies \cite{cartoixa1} have
confirmed this. Nevertheless all these studies rely on introducing terms
in the $\vec{k}\cdot\vec{p}$ model that mimic the symmetries of the structure
under consideration and have thus to be tailored to particular situations.
In contrast any atomistic approach, like the Empirical Pseudopotential
Method, incorporates by construction the correct symmetry of the structure.
However no such calculations, or even experimental data, is, to our
knowledge, available for the case of spin splittings.
It should also be noted that in this particular case, zincblende
common-anion structures, no structural asymmetry is introduced and any
spin splitting cannot be attributed to the Rashba effect.

Given that linear or cubic terms exist in bulk it is expected that any band would
split linearly close enough to the $\Gamma$ point. The fact that the point group
of these structures with a common-anion $D_\mathrm{2d}$ allows these terms further reinforces
our belief that this must indeed happen. Nothing however tells us that the coefficient
associated with this phenomenon would be big enough to allow eventual technological
use of these structures. Nevertheless the case of spin splitting enhancement by
symmetric confinement that we have already encountered lets us believe that this
is the case.

\begin{figure}[H]
\begin{center}
\includegraphics[width=\mmin{10cm}{0.8\linewidth}]{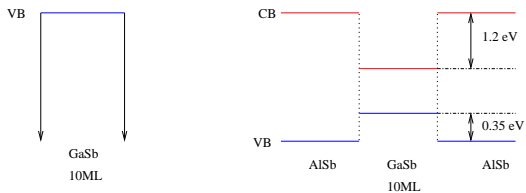}
\end{center}
\caption[Well layout for zincblende structures.]
{\label{layout3}Well layout for zincblende structures. The layout on the left
represents 10 monolayers of GaSb in infinite walls under the valence band while
the right layout represents a well of 10 monolayers of GaSb in AlSb.}
\end{figure}

\begin{figure}[H]
\begin{center}
\includegraphics[width=\mmin{12cm}{0.8\linewidth}]{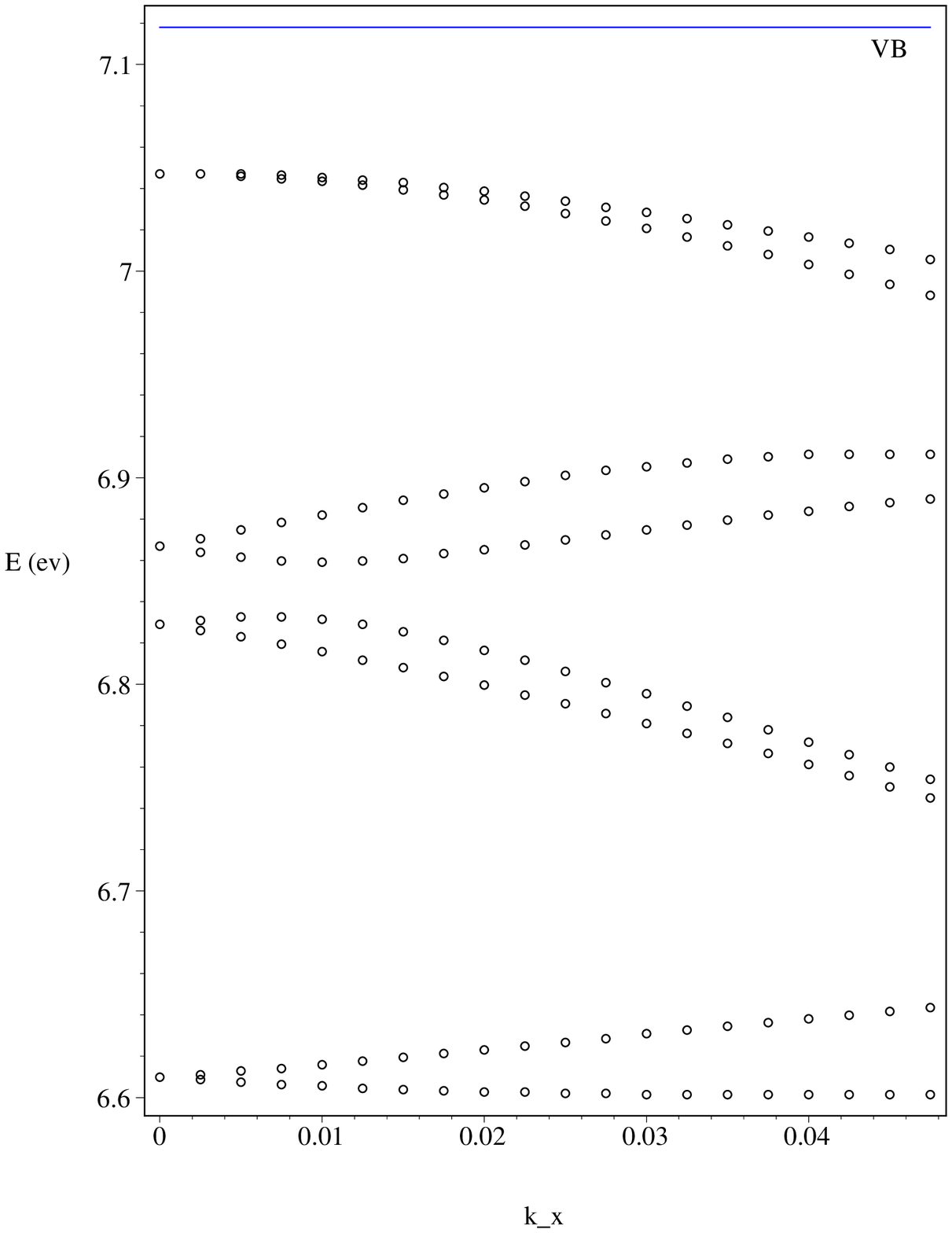}
\end{center}
\caption[Energy dispersion.]
{\label{well5}Energy dispersion for the layout represented on the left of
figure~\ref{layout3}. $\vec{k}_\parallel=(k_x,0)$ in units of
$\left(\frac{2\pi}{a}\right)$\AA$^{-1}$. VB represents the valence
band edge of GaSb.}
\end{figure}

A first introductory calculation with GaSb in infinite walls is then performed.
The band edge layout for this structure is depicted  on the left of figure
(\ref{layout3}). The energy dispersion computed with the infinite well model
is represented in figure~\ref{well5}.

Again we clearly see spin splittings that far exceed typical values of
bulk spin splittings. This is then another case of symmetric confinement
enhancement of the spin splittings. It should be noted that for this particular
direction there is no spin splitting in the bulk case.

A more realistic case of GaSb sandwiched between AlSb was also used. The
band layout is depicted on the right of figure~\ref{layout3}  where the
band offsets were set to acknowledged experimental
values.\cite{bandoffset}

\begin{figure}[H]
\begin{center}
\includegraphics[width=\mmin{12cm}{0.8\linewidth}]{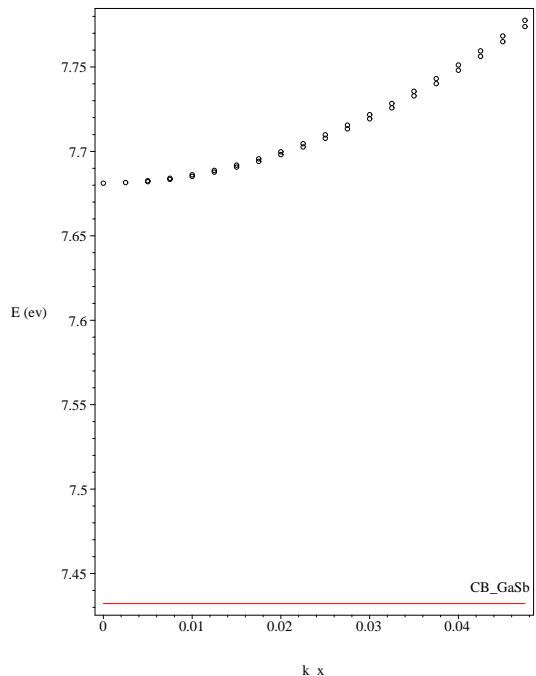}
\end{center}
\caption[Energy dispersion.]
{\label{well6}Energy dispersion for the layout represented on the right of
figure~\ref{layout3} for the energy window of conduction band.
$\vec{k}_\parallel=(k_x,0)$ in units of $\left(\frac{2\pi}{a}\right)$\AA$^{-1}$.
\mbox{CB\_GaSb} represents the conduction band edge of GaSb.}
\end{figure}

\begin{figure}[H]
\begin{center}
\includegraphics[width=\mmin{12cm}{0.8\linewidth}]{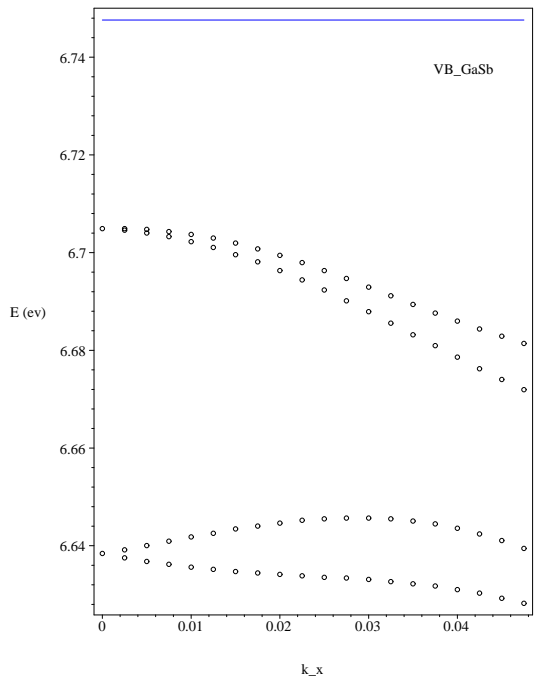}
\end{center}
\caption[Energy dispersion]
{\label{well7}Energy dispersion for the layout represented on the right of
figure~\ref{layout3} for the energy window of the valence band.
$\vec{k}_\parallel=(k_x,0)$ in units of $\left(\frac{2\pi}{a}\right)$\AA$^{-1}$.
\mbox{VB\_GaSb} represents the valence band edge of GaSb.}
\end{figure}

The computed energy dispersions for both the conduction and valence band
energy windows is shown in figures~\ref{well6} and \ref{well7} respectively.

Every band is clearly spin split: a fact that can be confirmed by a
calculation of the spin polarization. As before, for the [001] direction
no spin splitting is observed in the bulk case.  In both the conduction and
valence band the spin splitting can be easily fitted to a linear dispersion,
$\Delta E \propto k$, yielding a coefficient of $76.5\,\mbox{meV}\,\mbox{\AA}$
and $196.58\,\mbox{meV}\,\mbox{\AA}$ respectively.

%A representation
%of the actual spin splitting for the first band on the two energy windows is
%depicted on figures (\ref{spin1}) and (\ref{spin2}) respectively.
%As can be seen clearly the actual values of the spin splitting are not
%negligible and the linear coefficients, as calculated by a linear fit,
%are at least one order of magnitude larger than those typical for bulk.
%
%\begin{figure}[H]
%\begin{center}
%\includegraphics[width=\mmin{12cm}{0.8\linewidth}]{graphs/zincblende1.split.e.eps}
%\end{center}
%\caption[Spin splittings.]
%{\label{spin1}Spin splittings for the conduction level shown in
%figure (\ref{well6}).
%$\vec{k}_\parallel=(k_x,0)$ in units of $\left(\frac{2\pi}{a}\right)$\AA$^{-1}$.
%The red line represents a linear fit which produces a linear coefficient of
%76.5 meV\AA.}
%\end{figure}

%\begin{figure}[H]
%\begin{center}
%\includegraphics[width=\mmin{12cm}{0.8\linewidth}]{graphs/zincblende1.split.h.eps}
%\end{center}
%\caption[Spin splittings.]
%{\label{spin2}Spin splittings for the first valence level shown in
%figure (\ref{well7}).
%$\vec{k}_\parallel=(k_x,0)$ in units of $\left(\frac{2\pi}{a}\right)$\AA$^{-1}$.
%The red line represents a linear fit which produces a linear coefficient of
%196.58 meV\AA.}
%\end{figure}

\begin{figure}[H]
\begin{center}
\includegraphics[height=0.35\linewidth]{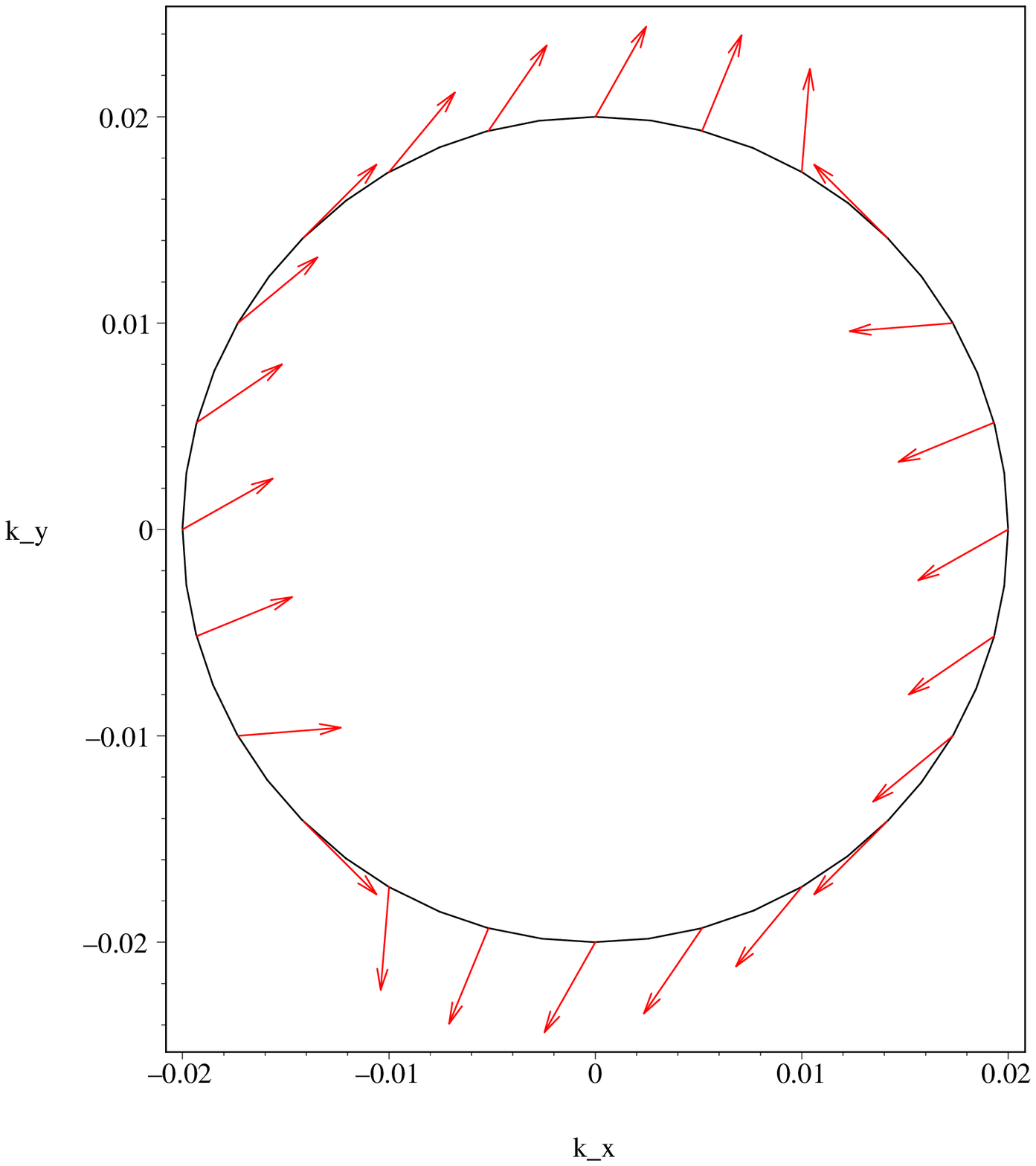}\hfil
\includegraphics[width=0.4\linewidth]{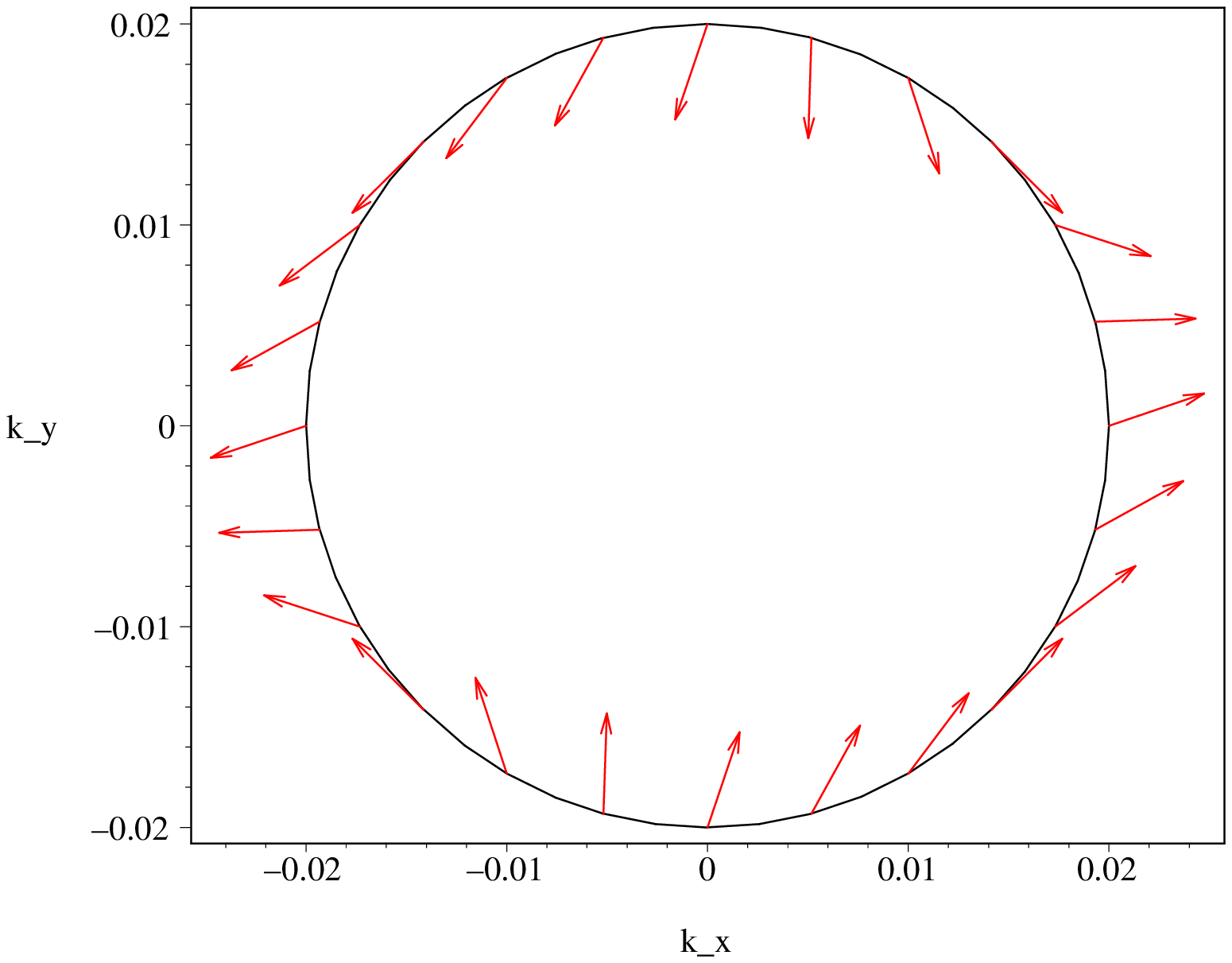}
\end{center}
\caption[Spin diagram.]
{\label{sdzb1}\label{sdzb2} Spin diagram for the first conduction level (left)
and valence level (right) in the band
structure of figures~\ref{well6} \& \ref{well7} respectively with
$k_\parallel=0.02\left(\frac{2\pi}{a}\right)$\AA$^{-1}$.
$\vec{k}_\parallel=(k_x,k_y)$ in units of
$\left(\frac{2\pi}{a}\right)$\AA$^{-1}$. Magnitude of spin scaled
for clarity.}
\end{figure}

The spin diagrams for a fixed
$k_\parallel=0.02\left(\frac{2\pi}{a}\right)$\AA$^{-1}$ and computed
for the first energy level on the energy windows for the conduction
and the valence bands are shown in figure~\ref{sdzb1}.
They represent a clear signature of the Dresselhaus terms and
cannot be attributed to any structural inversion asymmetry. The
slight deviation from the perfect $D_\mathrm{2d}$ signature was already
explained in section \ref{sym_sim}. In this case the spurious Rashba
term is approximately 0.34 times the Dresselhaus contribution. It
should be noted that the $z$ component of the spin polarization is always
found to be zero within numerical fluctuations. Spin diagrams for other
energy levels were computed with similar results.

It is important to note the values of the linear coefficients.
These values are comparable to the linear coefficients stated in
literature \cite{alistair} for the Rashba coefficient in
asymmetric structures. It should also be noted that these values cannot
be attributed to the spurious Rashba contributions in this case. The
situation with infinite walls does not suffer from this contamination
and produces similar results. This fact alone is technologically important: 
structural asymmetry is probably not
required to produce structures that behave similarly to those
currently proposed for the purpose of creating spin splittings.
The spin behavior as shown in the spin diagrams is completely
different, however.

As the SIA and BIA contributions have similar magnitudes it is important that
both are included in any study. This is particularly important when methods,
such as $\vec{k}\cdot\vec{p}$ are employed, which don't automatically contain
Dresselhaus contributions.  Even when the Rashba contribution is important the
interplay of the 2 terms may produce sizable effects.

\begin{figure}[H]
\begin{center}
\includegraphics[height=\mmin{12cm}{0.8\linewidth},width=\mmin{12cm}{0.8\linewidth}]{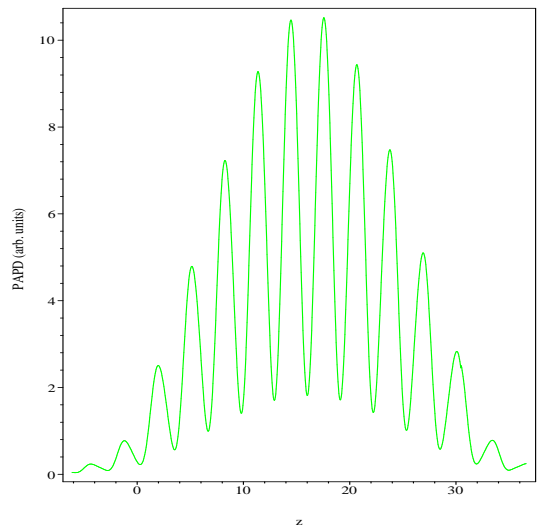}
\end{center}
\caption[Parallel averaged probability density.]
{\label{pdzb}Parallel averaged probability density for the first energy level
in the energy window of the valence band, as depicted in figure~\ref{well6}
for $\vec{k}_\parallel=(0.02,0)\left(\frac{2\pi}{a}\right)$\AA$^{-1}$.
Horizontal axis corresponds to the growth direction in {\AA}ngstr{\"o}m.}
\end{figure}

Another major achievement of this method is the possibility of extracting
atomistic details in clear contrast to the majority of the methods
previously used. In figure~\ref{pdzb} the parallel averaged probability
densities for the first energy level in the valence band energy window
is depicted.

Clearly the general trend is the same as in previous
calculations but a wealth of extra information is portrayed. The general
belief that the envelope behaves as predicted
by ``particle-in-box'' type calculations is confirmed even in this atomistic
calculation. Also displayed clearly is that there are deviations from the
envelope behavior in the atomistic detail of these graphs. 
This could help engineer particular structures tailored to
exhibit particular physical effects or even in the determination of the
best doping technique. Similar results are obtained for all other energy
levels.

\begin{figure}[H]
\begin{center}
\includegraphics[height=\mmin{4cm}{0.8\linewidth},width=\mmin{3cm}{0.8\linewidth}]{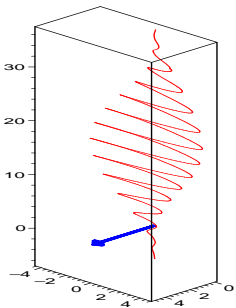}
\end{center}
\caption[Parallel averaged spin polarization.]
{\label{spzb}Three-dimensional representation of the parallel averaged spin polarization
$\vec{\sigma}(r_\perp)=(\overline{\sigma}_x,\overline{\sigma}_y,\overline{\sigma}_z)$,
corresponding to the parallel averaged probability density in figure~\ref{pdzb}.
Vertical axis corresponds to the growth direction in {\AA}ngstr{\"o}m. 
The continuous line corresponds to following the tip of the vector $\vec{\sigma}(r_\perp)$ in space after
appropriate scaling. The arrow is a vector in the direction of
$\vec{k}_\parallel$ introduced as guidance.}
\end{figure}

Atomistic detail is also clear in the parallel averaged spin polarization.
This is depicted in figure~\ref{spzb} corresponding to the parallel
averaged probability density depicted in figure~\ref{pdzb}.
This level of detail in spin behavior can be important in tailoring
particular spin properties and possibly in doping with magnetic materials.

It should be noted that most of the atomistic detail, in the particular
case of the Dresselhaus term, is in a plane perpendicular or nearly
perpendicular to $\vec{k}_\parallel$. It's however the component
parallel or nearly parallel to $\vec{k}_\parallel$ that averages
to the total spin polarization in accordance to the spin diagrams
characteristic of this term. This atomistic detail of the spin
polarization has never been reported previously.

It is also possible to use known data from bulk materials to compare
with our results. We know that for the conduction band spin splittings
are given by cubic terms originating in the Hamiltonian term given
by equation (\ref{cubicterms}). The constant $\gamma$ can be obtained
in the literature \cite{cardona3} and values range
from $109.4$ to $153.9\,\mbox{eV}\,\mbox{\AA}^3$ for theoretical predictions
with several methods and $186.3\,\mbox{eV}\,\mbox{\AA}^3$ for the experimental value. 
In the case of GaSb in AlSb the spin splittings of the levels in the
conduction band energy window should then originate, to first
approximation, in linear and quadratic terms in $k_{z,w}$, 
the value of the confined wavevector.
For wide enough wells this value should be, for the first energy
level, approximately $\frac{\pi}{L}$, with $L$ the well width given
by $\frac{1}{2}aN$ where $a$ is the lattice constant and $N$ the number
of monolayers in the well. 

\begin{figure}[H]
\begin{center}
\includegraphics[height=\mmin{12cm}{0.8\linewidth},width=\mmin{12cm}{0.8\linewidth}]{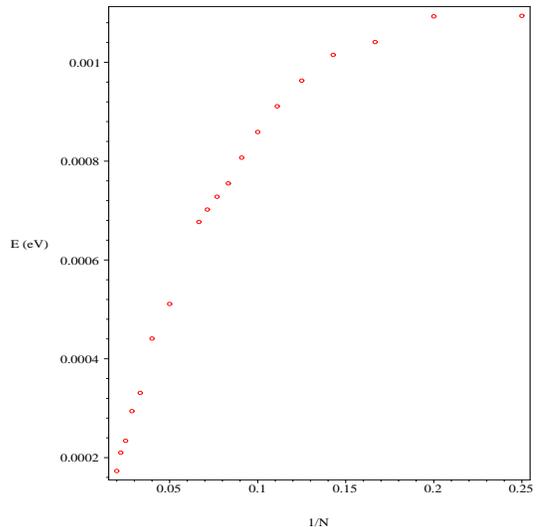}
\end{center}
\caption[Spin splitting well width dependence.]
{\label{wellwidthzb}Spin splitting of the first energy level in the
energy window of the conduction band for
$\vec{k}_\parallel=(0.01,0)\left(\frac{2\pi}{a}\right)$\AA$^{-1}$
as a function of the inverse well width given by the number of
monolayers of GaSb.}
\end{figure}

A well width dependence for the first electronic energy level was
then computed and the result is shown in figure~\ref{wellwidthzb}.

For the widest wells the linear contribution should
be dominant and the spin splitting should behave as:
\begin{equation}
\Delta=\frac{2\gamma\pi k_x^2}{a}\left(\frac{1}{N}\right).
\end{equation}
A linear fit
%, as shown in figure (\ref{wellwidthfitzb}), 
to this data gives a value of $\gamma$ of $118.2\,\mbox{eV}\,\mbox{\AA}^3$.

%\begin{figure}[H]
%\begin{center}
%\includegraphics[width=\mmin{12cm}{0.8\linewidth}]{graphs/wellwidth2.eps}
%\end{center}
%\caption[Spin splitting well width fit.]
%{\label{wellwidthfitzb}Spin splitting of the first energy level in the
%energy window of the conduction band for
%$\vec{k}_\parallel=(0.01,0)\left(\frac{2\pi}{a}\right)$\AA$^{-1}$
%as a function of the inverse well width given by the number of monolayers
%of GaSb in a restricted region. The red line corresponds to a linear fit
%to the data.}
%\end{figure}

Given all the approximations used this value is in very
good agreement not only with results using very different approaches,
but also with experiment, and gives us confidence in the reliability all of the
results of the method.

As a final conclusion on this section it should be noted that although
our results are for a particular case we believe that the dominant
trends in these structures are determined by symmetry and rather than by
the particular atoms involved. Further calculations for the case
of GaAs in AlAs were also computed and produced similar results.

%=======================================================

%%%%%%%%%%%%%%%%%%%%%%%%%%%%%%%%%%%%%%%%%%%%%%%%%%%%%%%%

%%%%%%%%%%%%%%%%%%%%%%%%%%%%%%%%%%%%%%%%%%%%%%%%%%%%%%%%
%
% Section VI:
%
%%%%%%%%%%%%%%%%%%%%%%%%%%%%%%%%%%%%%%%%%%%%%%%%%%%%%%%%

%=======================================================
                                         
\section{Asymmetric heterostructures}
                                                
%=======================================================

In the case of symmetric wells with a common anion we
probed the consequences of introducing the
Dresselhaus term. As a first approach we would be interested in
probing the Rashba contribution in the same way. The case
of asymmetric structures grown in the [001] direction is however
usually in the $C_\mathrm{2v}$ point group class which allows both terms.
However we already know that the Dresselhaus term arises from bulk
inversion asymmetry. Hence, if we consider a structure
with dominant structural inversion asymmetry we should minimize
the its effects. For this purpose we revisit the artificial structure 
of Ge sandwiched
between layers of Ge but with asymmetric band offsets as depicted
on the right of figure~\ref{layout2}. We should nevertheless remember
that, even though bulk Ge does not allow the Dresselhaus contribution,
this structure will contain this term even if it is small. We
have already shown that such contributions occur in cases where the
center of inversion has been removed as in the case of the
double layer of Si and Ge in infinite walls. However these
should be smaller than in any structure constructed out of
zincblende materials. As we shall see the Dresselhaus
contribution in this case is far smaller than the Rashba term.

The computed energy dispersion in the energy window of the valence
band was already displayed in figure~\ref{well4}. 
The bands clearly show spin splittings which have been confirmed
by a calculation of the spin polarization.
A linear fit to the actual spin splitting for the first band gives a splitting
coefficient of $324.3\,\mbox{meV}\,\mbox{\AA}$.

%\begin{figure}[H]
%\begin{center}
%\includegraphics[width=\mmin{12cm}{0.6\linewidth}]{graphs/asymmetric1.split.h1.eps}
%\end{center}
%\caption[Spin splittings.]
%{\label{ssazb}Spin splitting for the first valence level shown in figure
%(\ref{well4}).
%$\vec{k}_\parallel=(0.01,0)\left(\frac{2\pi}{a}\right)$\AA$^{-1}$.
%The red line represents a linear fit which produces a linear coefficient
%of 324.3 meV\AA.}
%\end{figure}

This value is of the same order as those calculated for the case of symmetric
structures;
further reinforcing our conclusion that BIA must always be taken into
consideration.

\begin{figure}[H]
\begin{center}
\includegraphics[width=\mmin{3.5cm}{0.5\linewidth}]{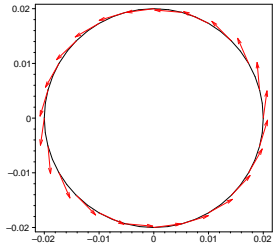}
\end{center}
\caption[Spin diagram.]
{\label{sdazb}Spin diagram for the first valence level with
$k_\parallel=0.02\left(\frac{2\pi}{a}\right)$\AA$^{-1}$.
$\vec{k}_\parallel=(k_x,k_y)$ in units of
$\left(\frac{2\pi}{a}\right)$\AA$^{-1}$. Magnitude of spin scaled
for clarity.}
\end{figure}

More importantly we have also calculated the spin diagram for this case
which is given in figure~\ref{sdazb}. In this case the $z$-component of
the spin polarization is also found to be zero within numerical fluctuations.
As we can see it forms a clear signature of the Rashba contribution.
The slight deviation is due, in this case, to the Dresselhaus term
which is allowed by the symmetry arrangement. This term
is approximately $0.14$ times the dominant Rashba contribution.

\begin{figure}[H]
\begin{center}
\includegraphics[height=\mmin{12cm}{0.8\linewidth},width=\mmin{12cm}{0.8\linewidth}]{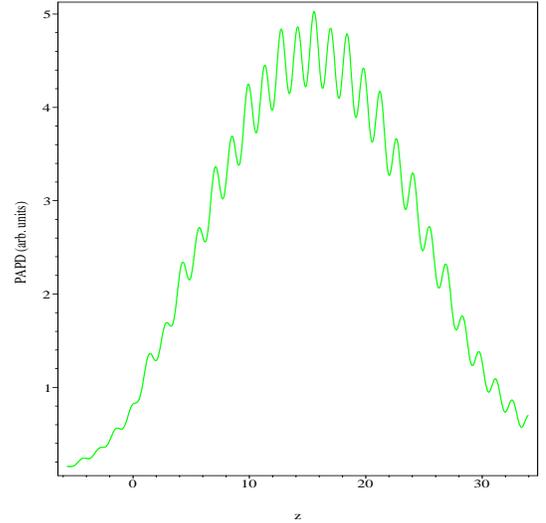}
\end{center}
\caption[Parallel averaged probability density.]
{\label{pdazb}Parallel averaged probability density for the first energy level
in the energy window of the valence band, as depicted in figure~\ref{well4},
for $\vec{k}_\parallel=(0.02,0)\left(\frac{2\pi}{a}\right)$\AA$^{-1}$.
Horizontal axis corresponds to the growth direction in {\AA}ngstr{\"o}m.}
\end{figure}

In this situation it is also possible to extract atomistic details
from our results. In figure~\ref{pdazb} a parallel averaged
probability density is shown.

\begin{figure}[H]
\begin{center}
\includegraphics[height=\mmin{4cm}{0.8\linewidth},width=\mmin{3cm}{0.8\linewidth}]{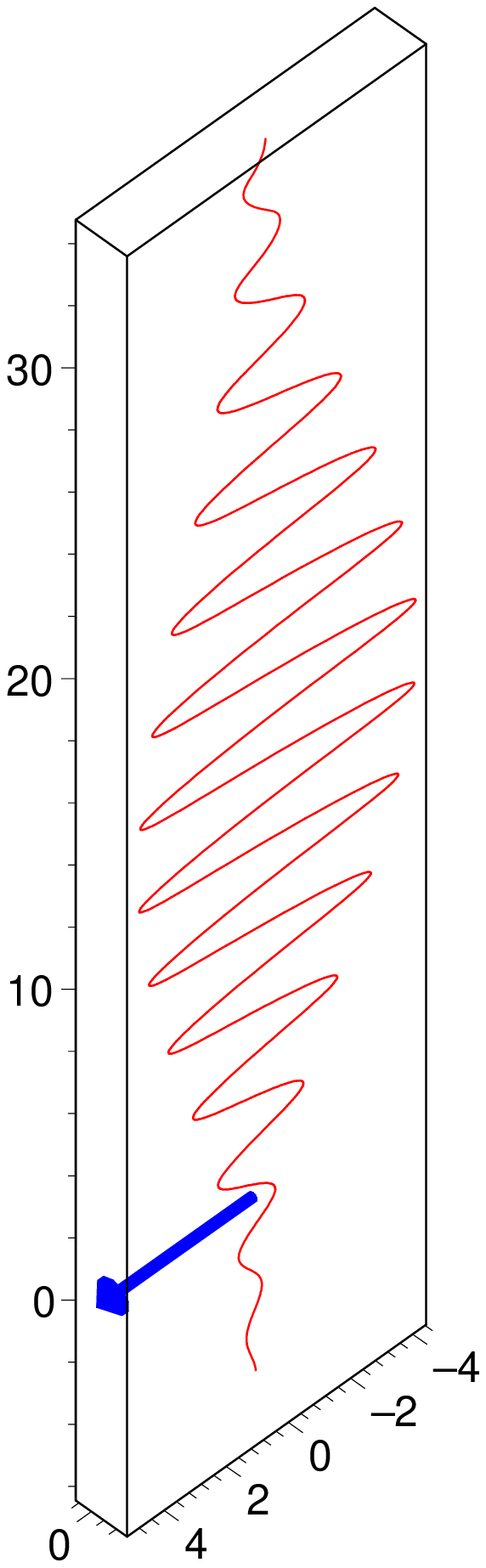}
\end{center}
\caption[Parallel averaged spin polarization.]
{\label{spazb}Three-dimensional representation of the parallel averaged spin polarization
$\vec{\sigma}(r_\perp)=(\overline{\sigma}_x,\overline{\sigma}_y,\overline{\sigma}_z)$,
corresponding to the parallel averaged probability density in figure~\ref{pdazb}.
Vertical axis corresponds to the growth direction in {\AA}ngstr{\"o}m. The continous line
corresponds to following the tip of the vector $\vec{\sigma}(r_\perp)$ in space after
appropriate scaling. The arrow is a vector in the direction of
$\vec{k}_\parallel$ introduced as guidance.}
\end{figure}

The typical envelope behavior is reproduced but significantly more
information is present. As in the previous case this
might be significant for engineering new structures. Atomistic
detail is also present in the parallel averaged spin
polarizations. This quantity is depicted in figure~\ref{spazb}
corresponding to the parallel averaged probability density shown
previously.

This case is different from the case of the Dresselhaus term.
Most of the atomistic detail is in the direction of $\vec{k}_\parallel$
while the perpendicular component averages to the total
spin polarization in accordance to the spin diagrams characteristic
of the Rashba term. Again this level of atomistic detail has never
been reported previously.

The main conclusion of our calculations is that structural asymmetry
in the system introduces a new term in the Hamiltonian that allows
linear splittings close to the $\Gamma$ point in perfect agreement
with the Rashba effect. It is also important to emphasis that the
Empirical Pseudopotential Method is capable of simulating this
situation.

It is possible to minimize the BIA contribution
by using diamond-like materials rather than producing structures with
point group $C_\mathrm{4v}$. However in a general case both terms will
interplay and the magnitude of the Dresselhaus contribution can be
comparable to that of the Rashba term. A detailed treatment of these
structures must always incorporate both to guarantee an
accurate description of the physical processes. This was previously 
discussed in the literature \cite{cartoixa1} using
the independent $\vec{k}\cdot\vec{p}$ method.

%=======================================================

%%%%%%%%%%%%%%%%%%%%%%%%%%%%%%%%%%%%%%%%%%%%%%%%%%%%%%%%

%%%%%%%%%%%%%%%%%%%%%%%%%%%%%%%%%%%%%%%%%%%%%%%%%%%%%%%%
%
% Section VII:
%
%%%%%%%%%%%%%%%%%%%%%%%%%%%%%%%%%%%%%%%%%%%%%%%%%%%%%%%%

%=======================================================
                                         
\section{Conclusions\label{concl}}
                                                
%=======================================================

We have shown that the phenomenon
of spin splittings in heterostructures can be put into a consistent
global framework. Until now the physics of symmetric and asymmetric
structures was thought to be in essence different. We have proven that
they are however just different expressions of the same underlying
physics: ``symmetry rules".

The Empirical Pseudopotential Layer Method
is very well suited for atomistic detailed calculations for these
structures. It can accurately predict the energy levels and their
spin splittings throughout the Brillouin zone. The case of the
neighborhood of the $\Gamma$ point, specifically analyzed in this chapter due to
its technological importance, is just a particular case.
The method can also play a crucial role in determining both
transport and optical properties of these structures as the
wavefunction for every possible state is easily obtained.
It is also flexible enough to handle arbitrary growth
directions and automatically incorporating the correct symmetries,
in contrast to the $\vec{k}\cdot\vec{p}$ method which requires to
be adapted for each particular symmetry case.

Our most significant conclusion
is that in all the studied structures linear terms emerge in the
spin splittings. These terms play a crucial part in determining
the top or bottom of bands near the $\Gamma$ point and are usually
the determining factor in experimental results that probe close
to that point.

The results involving diamond-like materials also provide substantial
conclusions. Although bulk materials with this structure do not exhibit
spin splittings we proved that heterostructures with these materials may
exhibit them. This is of crucial importance as most of today's semiconductor
industry is based on Si and Ge. The technology is still not capable
of producing structures with absolute layer precision and so the case of
odd number of atomic layers is probably unlikely to be of use. However
this apparent technological problem can actually be explored to produce
interface roughness that is responsible for lowering of symmetry;
thus making spin splittings possible \cite{golub}.

Symmetric structures with a common anion exhibit linear splittings comparable
to those determined in asymmetric
structures but with different spin behavior. The new field
of spintronics may find this new degree of freedom technologically useful.
The combination of both Dresselhaus and Rashba may be used to modulate a
particular spin behavior.

Our results are in good agreement with the most recent theoretical
study \cite{cartoixa1} in the literature. All linear coefficients for both
valence and conduction bands are similar even though the methods used
are different. The conclusion that both BIA and SIA must always be included
in every calculation is also drawn there.

Comparison with experiment is more difficult, however. Firstly there is
a lack of results for the diamond-like structures and for the
zincblende symmetric common-anion cases due to the relevance always given
to the asymmetric case. Secondly, even for the case
of asymmetric structures, it has emerged recently that the most common
experimental procedure, using Shubnikov-de Haas oscillations, might not
have been properly analyzed \cite{winkler1}. Neveretheless the
values obtained \cite{alistair} are of the same order of magnitude.

There is also another indirect experimental result that can be related
to our prediction.
An in-plane polarization anisotropy is observed \cite{murdin}
in the case of structures of $C_{2v}$ point group relevant to optical
considerations involving interband transitions \cite{magri}.
This anisotropy reveals itself in the band structure of these heterostructures
by a clear difference between the [110] and the [$\overline{1}$10] directions.
This difference is visible in our computed band structures for the asymmetric
case. For the common-anion symmetric situation no appreciable difference
could be detected. This effect is clearly important for optical devices
and can also be determined by the Empirical Pseudopotential Method.

Another point usually considered \cite{cartoixa1} in the context of spin
splittings in heterostructures is the influence of the main gap on the relative
magnitude of both contributions. This claim states that in narrow gap systems
SIA effects dominate while for wide band gaps BIA is larger.
Although we do not have enough data to compare linear coefficients for narrow
and large band gaps our results are enough to conclude that even in narrow gap
systems, like GaSb, both contributions should be taken into consideration.
We consider this claim to be an oversimplification of the dependence of
spin splittings on the parameters of the structure. These will depend in a
non-simple way not only on the main gap but also on the well width, depth and
applied fields. Results obtained \cite{cartoixa1} elsewhere seem to
confirm this.

The results presented here are the first, to our knowledge, atomistic
simulations that show that the full spin-orbit interaction caused by
the atomic cores is the dominant contribution for the zero-field
spin splittings. The particular symmetry of the case under
consideration determines the possible behavior:
Dresselhaus or Rashba.

%=======================================================

%%%%%%%%%%%%%%%%%%%%%%%%%%%%%%%%%%%%%%%%%%%%%%%%%%%%%%%%

\bibliography{biblio/biblio1,biblio/biblio2,biblio/biblio3,biblio/biblio4,biblio/biblio5,biblio/biblio6}

\end{document}